\documentclass[twocolumn,aps,prl,superscriptaddress,twoside,showpacs]{revtex4}
\pdfoutput=1
\usepackage{graphicx}
\usepackage{latexsym}
\usepackage{mathptmx,courier,helvet}
\graphicspath{{pic/}}
\usepackage[colorlinks=true,citecolor=blue,bookmarks=false]{hyperref}

\begin{document}

\title[organic heterojunction]{Charge carrier induced barrier height
         reduction at organic heterojunctions }

\author{S. W. Tsang}
\affiliation{%
Department of Material Science and Engineering, University of Toronto,
Toronto, Ontario, Canada M5S 3E4}

\author{M. W. Denhoff}
\affiliation{%
Institute for Microstructural Sciences, National Research Council,
Ottawa, Canada K1A 0R6}%% with \\

\author{Y. Tao}
\affiliation{%
Institute for Microstructural Sciences, National Research Council,
Ottawa, Canada K1A 0R6}

\author{Z. H. Lu}
\email{zhenghong.lu@utoronto.ca}
\affiliation{%
Department of Material Science and Engineering, University of Toronto,
Toronto, Ontario, Canada M5S 3E4}

\date{August 2008, to be published in Physical Review B (Vol.78, No.8),
      http://link.aps.org/abstract/PRB/v78/e081301}

\begin{abstract}
In order to provide an accurate theoretical description of current
density  voltage $(J-V)$ characteristics of an organic heterojunction
device over a wide range of electric fields at various temperatures,
it is proposed that an accumulation of charge carriers
at the heterojunction will lead to a reduction in the barrier
height across the heterojunction. Two well-known hole transporting
materials, 4,4,4-Tris(N-3-methylphenyl-N-phenyl-amino)
triphenylamine (MTDATA) and
N,N-diphenyl-N,N-bis(1-naphthyl)(1,1-biphenyl)-4,4diamine (NPB) were
used to fabricate unipolar heterojunction devices.
It is found that the $J-V$ characteristics depends strongly on applied bias.
The simulated $J-V$ characteristics of the heterojunction device,
with the modified injection model,
are found to be in
excellent agreement with the experimental data.
%Valid PACS numbers may be entered using the \verb+\pacs{#1}+ command.
\end{abstract}

%\pacs{41.20.Cv, 85.30.De}% PACS, the Physics and Astronomy 
%41.20.Cv  Electrostatics; Poison and Laplace equations, 
%               boundary value problems
% 85.30.De  Semiconductor-device characterization, design, and  modeling

\maketitle
\thispagestyle{plain}

Heterojunctions formed by two semiconductors has been the foundational 
technology for many modern electronic devices since 1950s~\cite{shockley}. 
The first high efficiency organic light-emitting-diode (OLED) structure 
based on a stack of multi-layer organic materials~\cite{tang1} has become a 
standard device platform in commercial production. Recently,  
5\%  power conversion efficiency from organic photovoltaic (OPV) 
cells has been demonstrated in small-molecule heterojunction devices 
and in polymeric blend structures~\cite{xue,peet}, and thus low cost OPV cells 
with current printing technology become commercially attractive. 
Despite those technological achievements with the applications of 
organic heterojunctions, the studies and understanding of the physics 
of the charge injection process across organic heterojunction are 
still very limited. 

There are two key factors that control the injection current at a 
heterojunction. One is the energy level alignment which determines 
the barrier height. According to recently reported photoemission studies, 
the interface alignment between two undoped organic materials 
often agrees with the vacuum level alignment rule. However, when 
one side of the heterojunction has extrinsic carriers, introduced 
either by chemical doping or electrical doping due to charge carrier 
accumulation, a considerable re-alignment of energy levels at the 
interface has been observed~\cite{vazquez, kahn, tang2, braun}. 
We have also reported that the injection current across an organic 
heterojunction device can be tuned by inserting a thin 2 nm 
chemically-doped interlayer with different doping concentrations 
at the interface~\cite{tsang1}. 
%
%This suggests that a second energy reference level, the quasi-Fermi 
%level, has to be considered as is the case for conventional 
%semiconductor heterojunctions. 
%
The other factor is the nature of the charge 
injection process. It is generally believed that the charge transport 
in organic materials is governed by the Miller-Abrahams type hopping 
process~\cite{miller} between localized transporting sites. 
Recently, Arkhipov \textit{et al}~\cite{arkhipov1} have proposed a two step hopping 
model to describe the charge carrier transport across an interface 
of disordered organic dielectric. The injection model considers the 
jump rate of a carrier from the center of the Gaussian density-of-state 
(DOS) in the injecting layer into the DOS of the accepting layer. 
However, the model does not include the energy distribution and 
occupation of carrier sites at the interface. 
Woudenbergh \textit{et al}~\cite{woudenbergh} found that measurements of a polymeric 
heterojunction device do not match the injection model. 
They modified the model to account for the charge filling effect 
of holes in the injecting layer by assuming the carriers are 
injected from the quasi-Fermi level rather than from the center 
of the highest-occupied-molecular orbital (HOMO).

In this letter, systematic experimental current density  
voltage $(J-V)$ characteristics of organic heterojunction devices made 
from relatively well-understood organic molecules were studied. 
In order to describe the experimental results, we found that it is 
essential to include two critical parameters in the existing theory: 
a) energy distribution of carriers at the injecting layer interface, and 
b) dynamic barrier height, $\phi_{\mathrm{v}}$, associated with  
quasi-Fermi level shifting. 

The approach in this experiment is to use a unipolar device where 
the current is mainly determined by the charge injection process 
at the organic heterojunction. This requires that the contact 
resistance at the metal-organic (MO) interface for charge injection 
and the number of trap states in the bulk organic material are negligible. 
This can be achieved by using 4,4,4-Tris(N-3-methylphenyl-N-phenyl-amino) 
triphenylamine (MTDATA) and 
N,N-diphenyl-N,N-bis(1-naphthyl)(1,1-biphenyl)-4,4diamine (NPB) 
to fabricate the heterojunction device. 
MTDATA and NPB are prototypical hole-injecting and hole-transporting 
materials used in OLED technology. Due to the difference in HOMO 
levels of MTDATA (5.0-5.1 eV)~\cite{shirota} 
and NPB (5.4-5.6 eV)~\cite{kahn,tang2}, 
there exists an energy barrier of around 0.5 eV for hole injection 
at the MTDATA / NPB interface. 
It has also been demonstrated that an Ohmic contact is formed at the 
indium tin oxide (ITO) and MTDATA interface~\cite{tsang1,giebeler}. Moreover, the 
trap-free space-charge-limited conduction behavior in single layer 
devices supports that there are a negligible number of trap states 
in both materials~\cite{tsang1, giebeler, tsang2}.

A hole-only heterojunction device with a structure of 
anode / injecting layer / accepting layer / cathode was fabricated 
for this study. Pre-patterned ITO coated glass substrate was used 
as the anode. The substrate was cleaned by a sequence of ultrasonic 
solvent baths and followed by UV-ozone treatment. 
A 320 nm layer of MTDATA was thermally evaporated on top of ITO as 
the injecting layer. It was then covered by 390 nm of thermally 
evaporated NPB as the accepting layer. Finally, silver was evaporated 
on top of NPB to form the cathode. The organic and metal layers were 
fabricated in a single vacuum system with a base pressure of $10^{-8}$\,Torr.  
On the cathode side, a large energy difference between the
lowest  unoccupied molecular orbital (LUMO) of NPB (2.5 eV) and the
work  function of Ag (4.3\,eV) effectively restricts electron injection 
into the device.  The field and temperature dependent hole mobilities
of individual  organic material were characterized by the
time-of-flight (TOF) technique.  Details of TOF measurement has been
report elsewhere~\cite{tsang1}.  The measured mobilities were fitted to the
Gaussian-disordered  model (GDM)~\cite{bassler} to obtain the energetic disorder
of HOMO of $\sigma_\mathrm{MTDATA}  = 93$\,meV and 
$\sigma_\mathrm{NPB} = 90$\,meV.

\begin{figure}
\centering
\includegraphics[width=8cm]{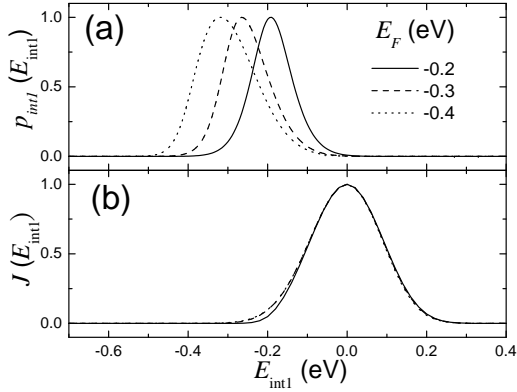} % for pdflatex
\caption{(a) The normalized energy distribution of the
carrier density, $p_{int1}(E_\mathrm{int1})$,  at the injecting layer
side of the interface.
(b) The normalized fractional injection current densities,
$J(E_\mathrm{int1})$  contributed
by the carriers at different energy levels. The results are compared
with different values of the quasi-Fermi level  (from $-0.2$\,eV to $-0.4$\,eV).}
\label{fig:1}
\end{figure}

In order to compare the injection model to the measured $J-V$ 
characteristics of an heterojunction device, the electric field  at the
interface, $F_\mathrm{int}$, and the charge carrier densities at both 
sides of the interface region, the injecting layer side,
$p_\mathrm{int1}$, and the accepting layer side, $p_\mathrm{int2}$,
have to be determined.  The complete two layer organic system including
contacts was modeled  using ATLAS~\cite{atlas}, to solve the standard electric
field, drift, and  diffusion equations used in semiconductor modeling.
The iteration  method was similar to the method used in Ref.~\cite{woudenbergh}. The
applied  voltages and measured current densities were used as the input
parameters to a steady-state model of the device. The field and 
temperature dependent mobilities measured by TOF were used in the 
simulation~\cite{tsang1}.  The hole-injecting electrode was modeled using a
Schottky contact  with the barrier height chosen to give a density of
holes equal to  $10^{21}$\,cm$^{-3}$ at the anode / injecting layer
interface. The cathode / accepting layer interface was specified as 
Ohmic (i.e. the hole concentration is pinned to the equilibrium  value
at this contact). ATLAS does not contain the hopping injection model.
Instead, we used a thermal injection model at  the organic
heterojunction and adjusted the parameters to give our  measured
current for each applied voltage.  This is equivalent to modeling the
two layers separately and using  the current and electric field as
boundary conditions at the interface. The results of the simulation
show that the holes pile up at the injecting layer side of the
interface creating a large electric field, which remains constant
across the accepting layer. In the range of applied voltage considered
here, there is a low and constant concentration of holes throughput the
accepting layer. The current in the accepting layer is simply given by
$J=e\mu p_\mathrm{int2}F_\mathrm{int}$.

\begin{figure}
\centering
\includegraphics[width=7cm]{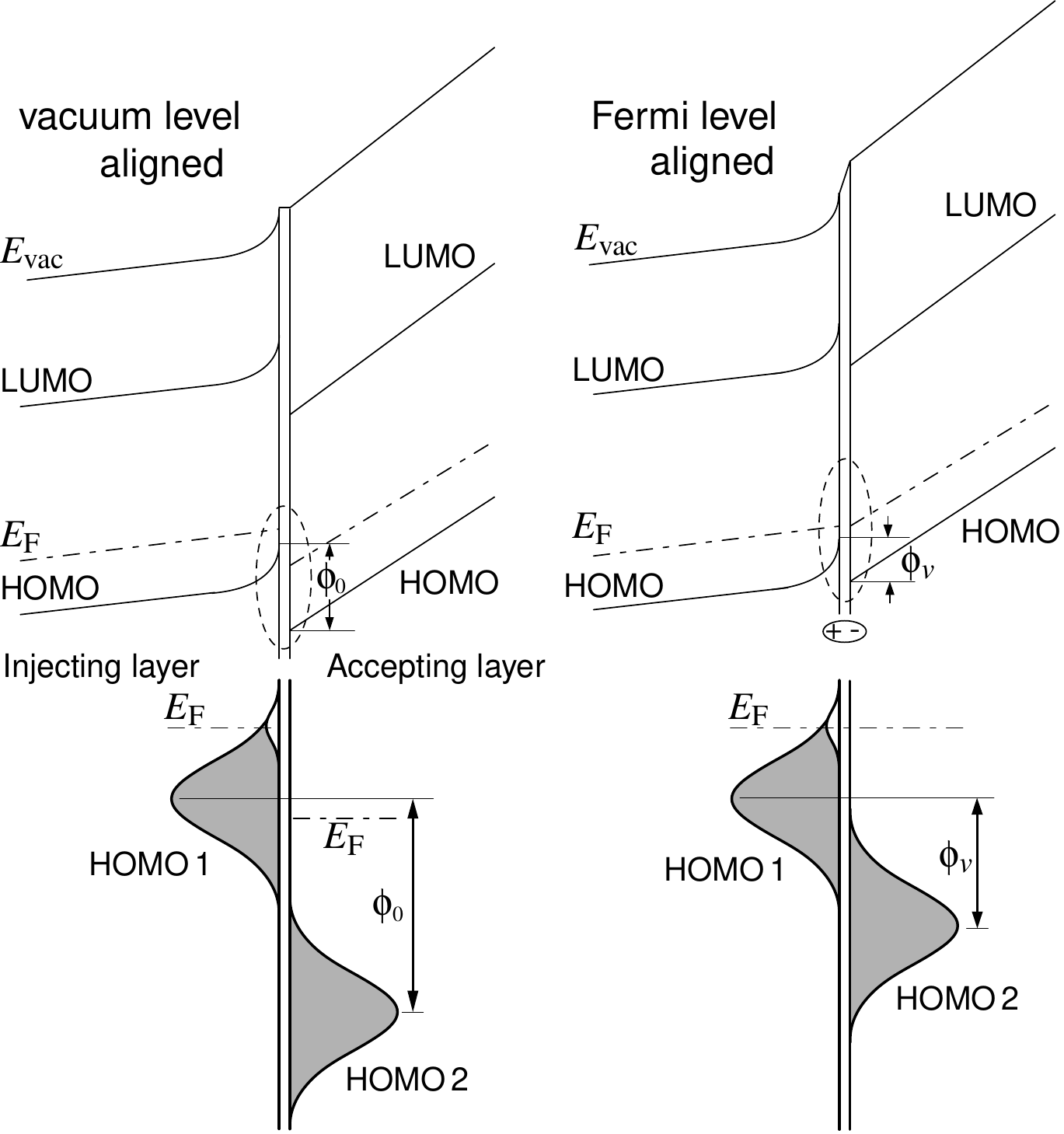}
\caption{A schematic diagrams of energy level alignment at the
heterojunction with the boundary condition of (a) vacuum level
alignment and (b) Fermi level alignment. The lower graphs are the
corresponding alignment of the Gaussian density-of-state (DOS) at both
sides of the interface. The shaded region represents those states
occupied by electrons.}
\label{fig:2}
\end{figure}

We have modified the injection model to account for the hole 
occupation by summing the contribution to the current of all the 
occupied hole states, $p_\mathrm{int1}(E)$,  at the injecting layer
side of the interface: 
\begin{eqnarray} 
J\  =\ \int_{-\infty}^{\infty}
\mathrm{d}E_{int1}\frac{p_\mathrm{int1}(E_\mathrm{int1})} {N_t}  \left[ e\nu_0
\int_a^{\infty} \mathrm{d}x\, \exp{(-2\gamma x)}\right. \nonumber\\
\times\left. \int_{-\infty}^\infty \mathrm{d}E_\mathrm{int2}
\mathrm{Bol}(\phi_0-E_\mathrm{int1}+E_\mathrm{int2}-eF_\mathrm{int}x) 
g(E_\mathrm{int2}) w_\mathrm{esc} \right]
\label{eq:J} 
\end{eqnarray} 
where $N_t$ is the total DOS, $e$ is the electron charge, $\nu_0$ is
the attempt-to-jump frequency of the order of the phonon frequency, 
$E_\mathrm{int1}$ is the energy of carriers at the injecting layer side
of the interface, $a$ is the nearest-neighbor distance, $\gamma$ is the
inverse localization radius, and $x$ is the hopping distance into the
accepting layer.  $E_\mathrm{int2}$ is the energy of the first site on
the accepting side,  $\phi_0$ is the barrier height, $F_\mathrm{int}$
is the interface electric field, 
%$g(E)=\frac{N_t}{\sqrt{2\pi} \sigma} \exp{(\frac{-E^2}{2\sigma^2})}$
$g(E)=N_t/\sqrt{2\pi} \sigma\ \exp{(-E^2/2\sigma^2})$ 
is the Gaussian
DOS in the accepting layer, and Bol$(E)$ is the energy dependence of
the jump rate: 
\begin{equation} 
\mathrm{Bol}(E) = \left\{ \begin{array}
             {r@{\quad \quad}l} 1\qquad\quad & E<0 \\ 
             \exp{(-E/kT)} & E>0
            \end{array} \right.  
\label{eq:Bol}
\end{equation}  
The expression inside the square
bracket in Eq.~(\ref{eq:J}), without the  term $E_\mathrm{int1}$, 
was  derived by Arkhipov
\textit{et al}~\cite{arkhipov1}. It calculates the  thermally assisted hopping rate for a
hole at the interface of the  injecting layer to a site over a
distance, $x$, into the accepting layer.  This is multiplied by the
escape probability, $w_\mathrm{esc}$, of the hole  continuing into the bulk of the
accepting layer (as opposed to  hopping back down the barrier). The
expression from Ref.~\cite{arkhipov1}  assumes that there is a hole, at the HOMO
level of the interface  of the injecting layer, for every receiving
site in the accepting layer.  Our modification accounts for the
occupancy and the energy distribution  of holes at the injecting layer
side of the interface.  In order to do this, the expression in the
square bracket is  multiplied by the probability that an injecting site
is occupied,  $p_\mathrm{int1}(E)/N_t$, and everything is integrated 
over $E_\mathrm{int1}$. 
The argument of $\mathrm{Bol}(E)$ was modified from 
$(\phi_0 + E_\mathrm{int2} - eF_\mathrm{int}x)$ in  Ref.~\cite{arkhipov1}
to $(\phi_0 - E_\mathrm{int1} + E_\mathrm{int2} - eF_\mathrm{int}x)$ 
to account for the energy 
distribution in the injecting layer.

Using the ATLAS simulation results for $p_\mathrm{int1}$, 
the quasi-Fermi level, $E_F$, and 
the hole occupation probability $p_\mathrm{int1}$  can be obtained using
\begin{equation}
p_\mathrm{int1}\!\!=\!\!\!\int_{-\infty}^\infty \!\!\!\!\!\!\mathrm{d}E 
p_\mathrm{int1}(E)\!\!=\!\!
 \frac{N_{t}}{\sqrt{2\pi}\sigma}
  \int_{-\infty}^\infty \!\!\!\!\!\!\mathrm{d}E
   \frac{\exp{(-\frac{E^2}{2\sigma^2})}}
    {\left[ 1+\exp{(\frac{E-E_F}{kT})} \right]}
\label{eq:pint1}
\end{equation}
where $\sigma$ is the width of the Gaussian DOS in the injecting layer.  
$p_\mathrm{int1}$ and the integrand of Eq.~(\ref{eq:J}), $J(E_\mathrm{int})$, 
are plotted as a function of $E_\mathrm{int}$
 in Fig.~\ref{fig:1} for different values of $E_F$. Remarkably, the current 
contributed by those carriers at $E_F$ is almost negligible. 
The main contribution is due to the carriers at the energy of the 
maximum of the DOS, even though the Fermi level and the majority of 
the carriers lie below this level. This demonstrates that the assumption 
made in Ref.~\cite{woudenbergh} that the carriers are injected from the Fermi 
level is incorrect. Mathematically, this peak is due to the exponentially 
falling factor of the Fermi function in $p_\mathrm{int1}(E)$ in 
Eq.~(\ref{eq:pint1}) canceling the 
exponentially rising factor in $Bol(E)$ in Eq.~(\ref{eq:Bol}), leaving the Gaussian 
in $p_\mathrm{int1}(E)$  as the predominant energy dependent factor. 

\begin{figure}
\centering
\includegraphics[width=7cm]{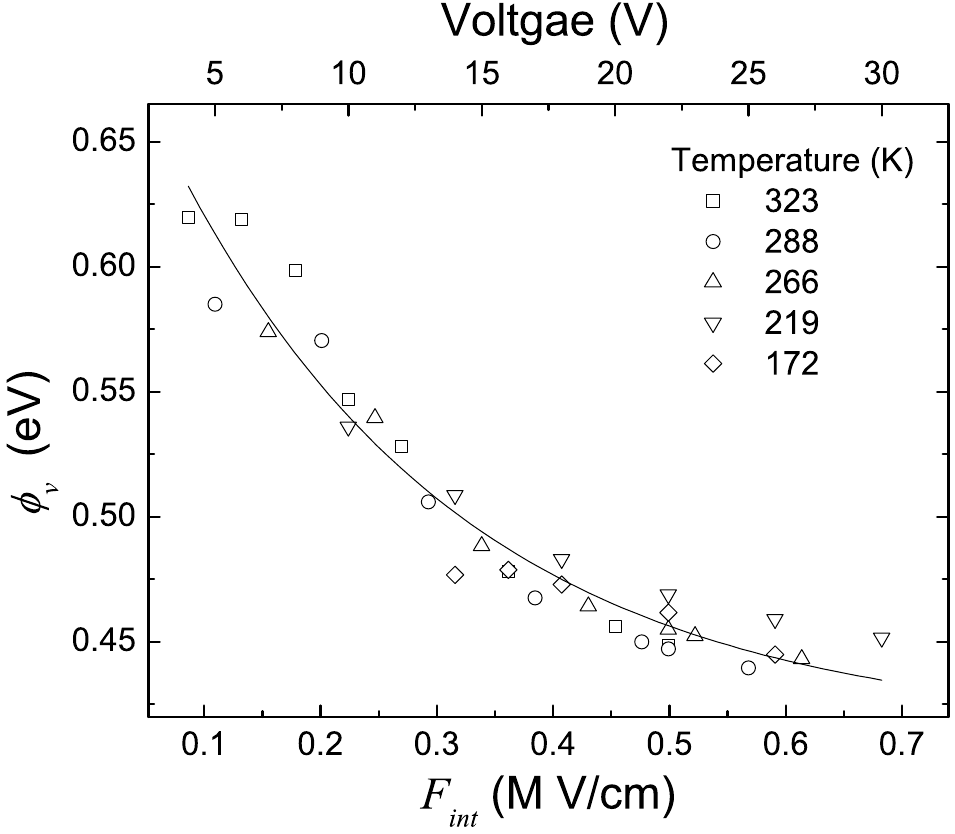}
\caption{The field and voltage dependence of the variable barrier $\phi_v$
at 323 K ($\Box$), 288 K ({\footnotesize$\bigcirc$}), 266 K ($\bigtriangleup$),
219 K ($\bigtriangledown$), and 172 K ($\Diamond$).
Assuming thermodynamic equilibrium, $\phi_v$ is calculated 
using Eq.~(\ref{eq:pint1})
with $p_\mathrm{int1}$ and $p_\mathrm{int2}$ obtained by the steady-state simulation.
The solid line is a fitting curve to an exponential decay function.}
\label{fig:3}
\end{figure}

The model described above assumes a constant energy level alignment 
at the organic heterojunction which is determined by imposing vacuum 
level alignment. However, the existence of an interface dipole would 
shift the vacuum level and change the alignment~\cite{ishii}.  
As discussed earlier in this letter, there is recent experimental 
and theoretical support for the existence of an interface dipole in, 
at least, some organic systems.  Considering this, we propose a 
model where the energy level alignment is determined by the changes 
in Fermi levels in the two organic layers.  The result of this is 
that a change in carrier concentrations will cause a change in the 
barrier height.  The simplest assumption is to assume thermodynamic 
equilibrium across the interface which implies that the Fermi level 
is continuous at the interface.

This model is illustrated in Fig.\ref{fig:2}, which shows the energy levels 
in the case with an applied voltage and a current in the device.  
The energy level bending in the injecting layer is due to the holes 
piling up at the interface.  Fig.~\ref{fig:2}(a) shows the case 
where the vacuum 
levels of the two layers are matched at the interface.  
In this case the Fermi level will be discontinuous.  
The barrier height, which is the difference between the HOMO 
levels at the interface, is constant.  Fig\ref{fig:2}(b) represents our model, 
where an interface dipole has formed which aligns the two Fermi levels.  
This will shift the vacuum levels and now the difference between the 
HOMO levels has shifted resulting in a smaller barrier height.  
Changing the current will change the hole concentrations, 
hence changing the Fermi levels. Evidently, this model features a 
variable barrier height $\phi_v$.

\begin{figure}
%\centering
\includegraphics[width=8cm]{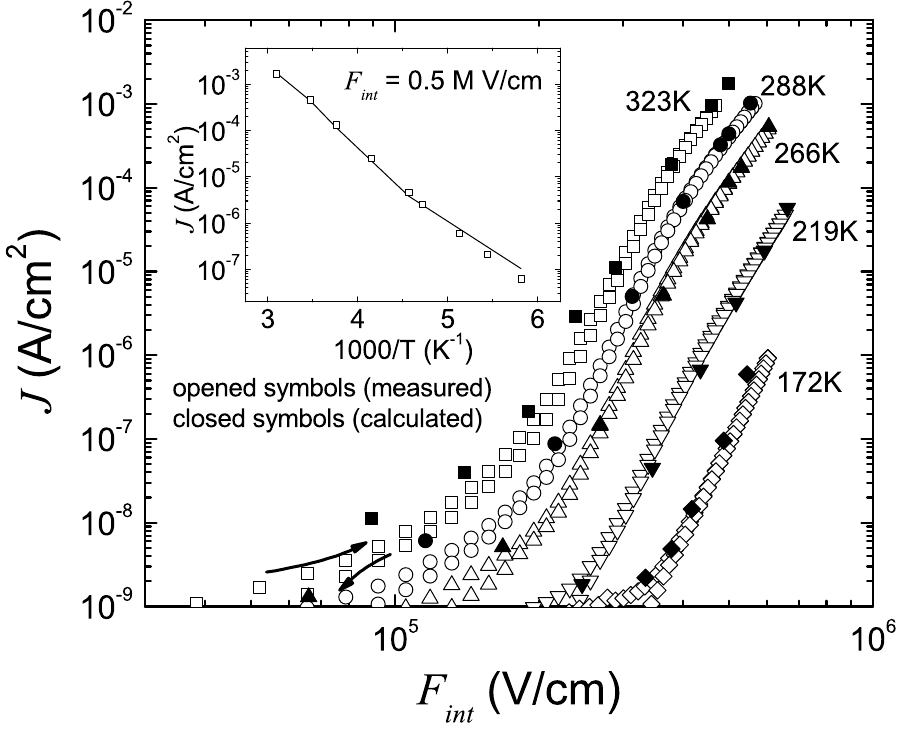}
\caption{Measured $J$ - $F_\mathrm{int}$ characteristics of the
ITO / MTDATA(320 nm) / NPB(390 nm) / Ag heterojunction device at
323 K ($\Box$), 288 K ({\LARGE$\circ$}), 266 K ($\bigtriangleup$),
219 K ($\bigtriangledown$), and 172 K ($\Diamond$).
The corresponding solid symbols are the calculation results based on
the modified injection model in Eq.~(\ref{eq:J}) with the variable barriers, $\phi_v$,
in Fig.~\ref{fig:3}. The inset is the $J$ vs $1/T$ plot at
$F_{int}$ = 0.5 M V/cm. Fitting the calculation results
(solid line) with Eq.~(\ref{eq:J}) to the experimental data 
(opened symbols) obtains $\phi_v = 0.45 \pm 0.1$\,eV,
$\sigma_{\mathrm{MTDATA}}  = 95$\,meV and
$\sigma_{\mathrm{NPB}} = 110$\,meV.}
\label{fig:4}
\end{figure}

Using Eq.~(\ref{eq:pint1}) and the simulated interface hole 
densities $p_\mathrm{int1}$ and $p_\mathrm{int2}$ of 
the ITO / MTDATA (320 nm) / NPB (390 nm) / Ag heterojunction device, 
the variable barrier, $\phi_v$ can be obtained by adjusting the 
energy alignment 
so that the Fermi levels $E_F$ match at the MTDATA/NPB interface. 
Our choice of the values of $\sigma_\mathrm{MTDATA}$ 
and $\sigma_\mathrm{NPB}$ used in the calculation will be 
discussed later. Figure~\ref{fig:3} shows the calculated $\phi_v$ as a function of 
applied voltage at different temperatures. $\phi_v$ varies little with 
temperature, but it decreases exponentially when the applied voltage 
increases. This can be explained by the occupancy of holes in the 
interface DOSs. According to the simulation results, with increasing 
voltage, $p_\mathrm{int1}$ increases from $10^{17}$\,cm$^{-3}$ to 
$10^{19}$\,cm$^{-3}$ and $p_\mathrm{int2}$ increases 
from $10^9$\,cm$^{-3}$ to $10^{13}$\,cm$^{-3}$. 
The amount of change of the Fermi levels depends on the changes in  
and $p_\mathrm{int1}$ and $p_\mathrm{int2}$ on the DOS 
(at the position of the Fermi levels). 
In this device, the Fermi level in the accepting layer changes more 
rapidly than that in the injecting layer, leading to a decreasing 
barrier as voltage increases. It can be deduced that, besides the 
energy difference of the HOMO levels, the energy profile is significant 
in controlling carrier injection across organic heterojunction.

The inset in Fig.~\ref{fig:4} illustrates the first step in our calculation 
for the ITO / MTDATA / NPB / Ag heterojunction device. An inverse 
localization radius $\gamma= 5\times 10^7\,\mathrm{cm}^{-1}$ 
has been used~\cite{wolf}. 
The nearest-neighbor distance $a = 1$\,nm is taken as the size of a 
NPB molecule. By assuming only one net charge is carried by a 
molecule at a given time, the total number of DOS 
$N_t = 1\times 10^{21}$\,cm$^{-3}$. 
For a fixed interface electric field ($F_\mathrm{int} = 0.5$\,MV/cm), 
the disorder parameters $\sigma_\mathrm{MTDATA}$ and $\sigma_\mathrm{NPB}$, 
and the attempt-to-jump frequency $\nu_0$, 
are varied to obtain the best fit to the temperature dependent data. 
Using the values $\nu_0 = 7.88\times 10^{13}$\,s$^{-1}$,
$\phi_v = 0.45 \pm 0.1$\,eV, $\sigma_\mathrm{MTDATA} = 95$\,meV 
and  $\sigma_\mathrm{NPB}= 110$\,meV, 
the theory agrees with the observed thermally 
activated current. 
%The interface Gaussian widths $\sigma_\mathrm{MTDATA}$ 
%and $\sigma_\mathrm{NPB}$ are in 
%good agreement with the bulk values obtained by the TOF measurements 
%and are used to calculate the $J$ vs $F_\mathrm{int}$ 
%characteristics of the device. 
The values for the disorder are slightly larger than those obtained 
from the TOF measurement.  
A broadening  would be expected in the presence of an interface 
dipole~\cite{baldo}.
The above values are used in Eq. (1), with the introduction of 
the variable barrier, $\phi_v$, to calculate the $J$ vs $F_\mathrm{int}$
characteristics in Fig.~\ref{fig:4}.
%As shown in Fig.~\ref{fig:4}, using Eq.~(\ref{eq:J}) with the 
%introduction of the variable barrier $\phi_v$, 
The calculated electric field and temperature 
dependent injection current densities are in excellent agreement 
with the experimental results. 
The good agreement between two sweeping directions of the applied 
voltage at all temperatures also suggests that the trap charging 
effect is negligible. It is worthwhile noting that ignoring the 
contribution of the variable barrier, $\phi_v$, will vastly underestimate the 
field dependence of the injection current across the heterojunction.

In conclusion, an injection model for an organic heterojunction which 
includes the distribution of carriers at the injecting side of the 
interface has been proposed. This model shows that the main contribution 
to the injection current is from carriers at the peak of the Gaussian 
density-of-state (DOS) rather than at the Fermi level. In addition, 
we have proposed a variable injection barrier height, $\phi_v$ which arises 
from the energy level re-alignment triggered by the relative Fermi 
level shifting caused by charge carrier accumulation at the interface. 
The model is verified by the excellent agreement with the 
experimental results of the current density  voltage $(J-V)$ 
characteristics of an organic heterojunction device for various 
temperatures over a wide range of electrical field. 
The results show that a dynamic change in carrier concentration 
and DOS energy alignment at the heterojunction plays a significant 
role in the performance of organic heterojunction devices.

\begin{acknowledgments}
Financial support for this project is provided by Ontario Centres of
Excellence (OCE), National Research Council Canada (NRC),
and Natural Science and Engineering Research Council of Canada (NSERC).
\end{acknowledgments}

%\section*{References}

\end{document}